\documentstyle[11pt,newpasp,twoside,epsf]{article}
\begin{document}

\title{The HI mass function in Ursa Major}
\author{Marc Verheijen}
\affil{NRAO, P.O. Box O, Socorro, NM 87801, U.S.A.}
\author{Neil Trentham}
\affil{Institute of Astronomy, Madingley Road, Cambridge CB3 0EZ, U.K.}
\author{Brent Tully}
\affil{Institute for Astronomy, 2680 Woodlawn Drive, Honolulu, HI 96822, U.S.A.}
\author{Martin Zwaan}
\affil{Kapteyn Institute, Postbus 800, 9700 AV, Groningen, The Netherlands}

\begin{abstract}
 A deep blind HI survey of the nearby spiral rich Ursa Major cluster has
been performed with the VLA in parallel with a wide-field R-band imaging
campaign of most VLA fields with the CFHT. The goal is to measure the
slope of the HI mass function (HIMF) down to HI masses of
10$^7$M$_\odot$ as well as the slope of the faint-end of the
luminosity~function (LF) down to M(R)=$-10$. The VLA survey sampled 16\%
of the Ursa Major cluster volume. Both the HIMF and the LF in Ursa Major
are nearly flat at their faint-ends. No free-floating HI clouds were
detected.
 \end{abstract}

\section{Motivation}

\begin{figure}
\plotone{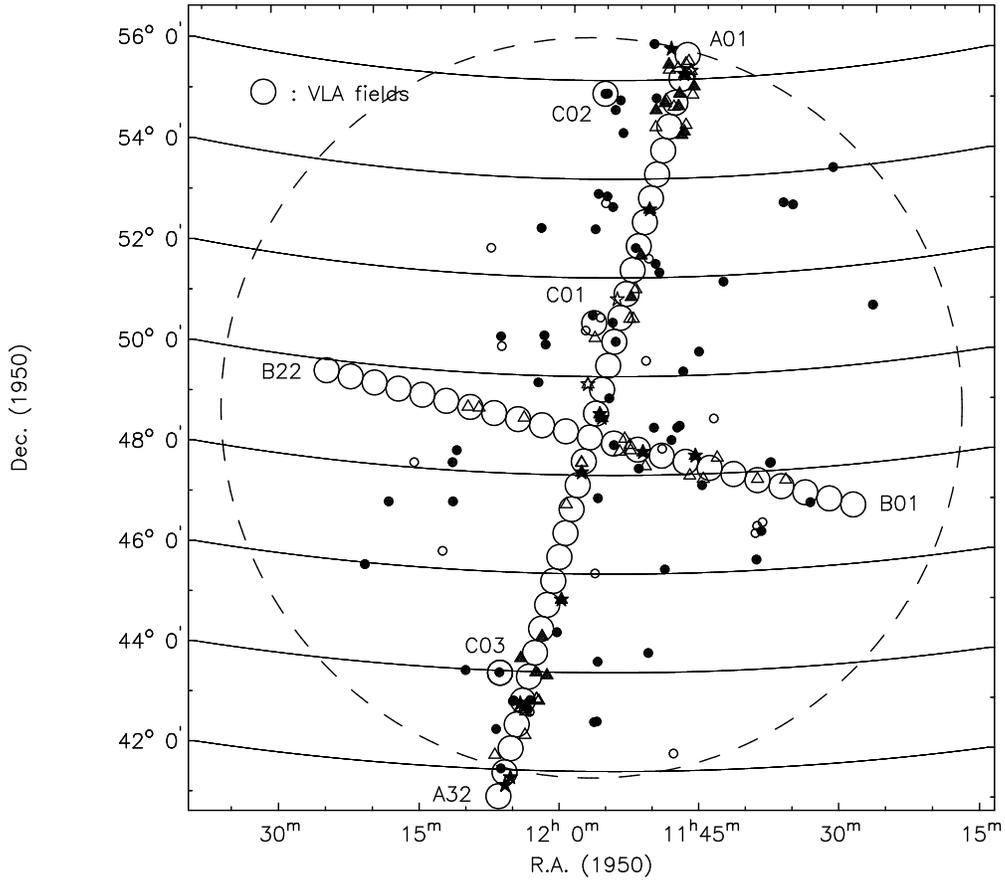}
 \caption{The layout of the VLA pointings in the Ursa Major cluster.
Note that both crowded and empty regions in the cluster are sampled.
{\bf Circles}: previously known cluster members (solid: complete sample,
open: galaxies intrinsically fainter than the SMC). {\bf Stars}: Newly
discovered HI dwarfs (solid: inside cluster velocity range, open:
outside velocity range). {\bf Triangles}: optically selected cluster
candidates( solid: almost certain member, open: probably background
objects). The C-fields were pointed observations toward interesting
galaxies and were omitted from the HIMF analysis. The large circles
indicate the size of the FWHM of the VLA primary beam.}
 \end{figure}

The slope of the HI mass function (HIMF) at the low mass end is a much
debated topic. It is related to the distribution of primordial density
fluctuations and it helps constraining galaxy formation scenarios. In
case a large undiscovered population of HI rich dwarf galaxies does
exist, it may contain a non-negligible fraction of the baryon content of
the universe and provide fuel for the star formation process in more
massive disk galaxies through their capture, infall and accretion.
Furthermore, the slope of the HIMF ties in directly with the recently
launched hypothesis that the population of High Velocity Clouds (HVCs)
as cataloged by Wakker \& Van Woerden (1991), apparently associated with
the Milky Way, might actually be of an extragalactic nature, being the
primordial remnants of the formation of the Local Group (Blitz et al.
1999). 

Despite its importance, the slope of the HIMF at the low-mass end has
not been measured with any accuracy down to HI masses of
10$^7$M$_\odot$. In general, the statistics in the $10^8-10^7$M$_\odot$
mass range are too poor, with typically 1 or 2 galaxies in the lowest HI
mass bin. The Arecibo HI Strip Survey (AHISS) of Zwaan et al. (1997),
still the deepest survey to date, indicates a slope in the range $-$1.1
to $-$1.3 with only 1 galaxy in the last bin. From the AHISS data,
combined with an independent deep Arecibo Slice Survey, Schneider et al.
(1998) claimed a sharp steepening of the HIMF below 10$^8$M$_\odot$ with
only 2 galaxies in the last bin. A first HIMF comprising 263 galaxies
from the HIPASS project (Kilborn et al. 1999) shows a slope of $-$1.3
with only 3 galaxies in two bins below an HI mass of 10$^8$M$_\odot$. At
this meeting in Guanajuato, an updated slope of roughly $-$1.4 was
reported from the HIPASS survey by Staveley-Smith and a slope as steep
as $-$1.7 was suggested by Schneider from the new Arecibo Dual-Beam
Survey (ADBS). It should be noted that the volume corrections of all
these surveys are hampered by uncertain corrections for large scale
structure in the Universe. These corrections should account for the fact
that we are located in a local overdensity and that dwarf galaxies are
more easily detected nearby than in the distant Universe, introducing a
bias toward a steeper slope.

From all this one can conclude that the sensitivity of recent surveys is
simply insufficient and deep observations of a well defined,
significantly overdense region with field-like characteristics of its
galaxy population are needed to obtain the required statistics. It was
realized that the sensitivity of the VLA in its D$-$configuration in
combination with the overdensity of the nearby spiral rich Ursa Major
(UMa) cluster would provide sufficient detections per HI mass-bin down
to 10$^7$M$_\odot$ to reliably measure the slope of the HI mass function
at the low-mass end. Furthermore, all galaxies in the UMa volume would
be at the same distance, eliminating the effects of large scale
structure along the depth of the survey. The VLA would also be sensitive
enough to detect the UMa equivalent of the proposed Local Group
population of extragalactic HVCs with typical HI masses of a few times
10$^7$M$_\odot$, as initially suggested by Blitz et al. (1999). 

\section{The Ursa Major cluster}

The Ursa Major cluster at a distance of 18 Mpc spans about 15 degrees on
the sky and comprises some 79 Mpc$^3$. It is a gravitationally bound
overdensity located in the Supergalactic plane. However, it contains no
ellipticals, just a dozen lenticulars and has a velocity dispersion of
only 150 km$\cdot$s$^{\mbox{\tiny -1}}$, which corresponds to a crossing
time of roughly a Hubble time (Tully et al. 1996). Its morphological mix
is similar to lower density field regions. This makes the Ursa Major
region an especially suitable environment to study the low-mass end of
the HIMF. For instance, if the slope of the HIMF would indeed be as
steep as $-$1.7 as suggested by Schneider in this meeting, there would
be about 2900 HI objects in the Ursa Major volume with HI masses above
10$^7$M$_\odot$, based on the high-mass end of the HIMF of the 80 known
cluster members.

\begin{figure}
\plotone{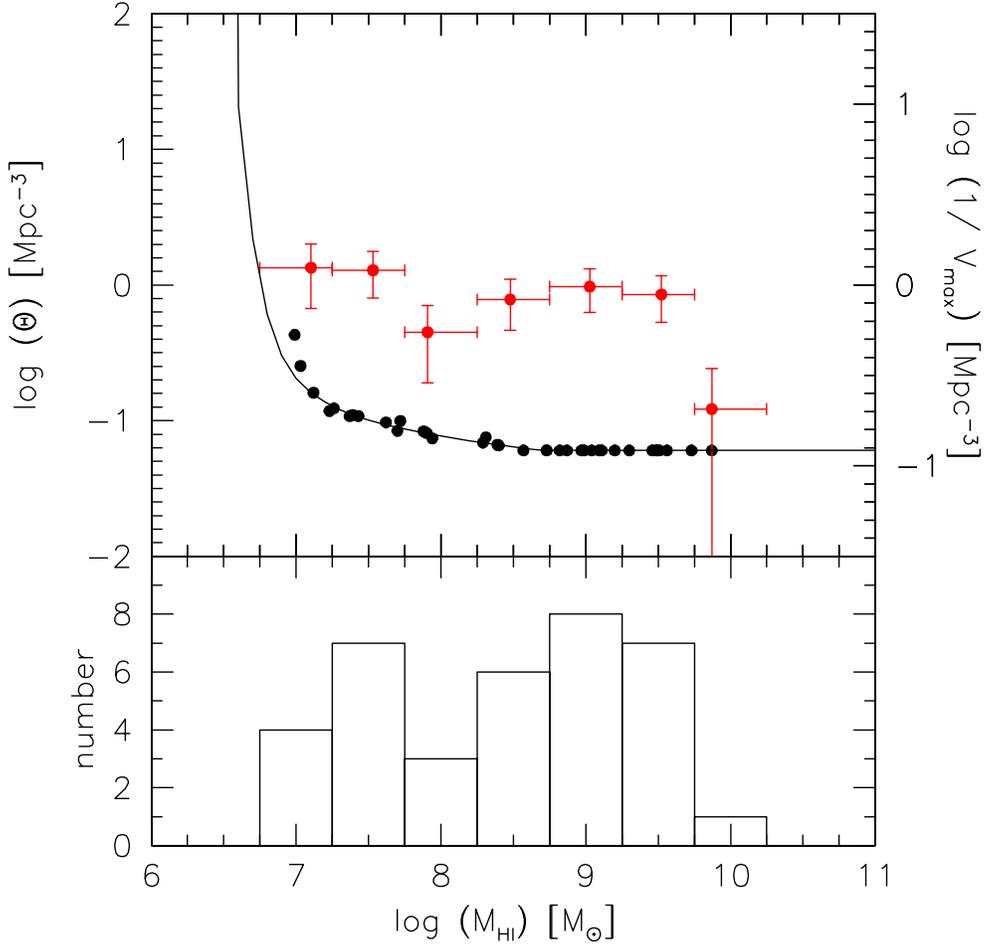}
 \caption{The HIMF in Ursa Major as measured with the VLA. {\bf Upper
panel}: Horizontal bars show the bin widths, vertical errorbars indicate
the Poisson noise. The positions of the points along the horizontal bars
indicate the average HI mass of the galaxies within each bin.  The solid
line shows the volume probed as a function of HI mass, assuming that the
detections follow an HI mass-line width relation 
$\Delta$V~$\propto$~M$_{\rm HI}^{1/3}$ and the profile shapes at the low
mass end are Gaussian. The dots indicate the actually applied 1/V$_{\rm
max}$ for individual galaxies, taking into account their measured HI
masses  and velocity widths. Their close adherence to the solid line
indicates the validity of the assumptions that go into the analytically
constructed volume corrections. {\bf Lower panel}: Number of detected
galaxies per bin.}
 \end{figure}

\section{The VLA blind HI survey}

A VLA blind HI survey of the Ursa Major cluster and CFHT imaging of most
of the VLA fields in the R-band with the wide-field camera has been
performed. The optical campaign was aimed at measuring the faint end of
the Luminosity Function (LF) while it would provide optical morphologies
and luminosities of the HI detected dwarfs. In return, the HI detections
could provide redshifts for the optically selected cluster candidates. 

A cross-pattern of 22$\times$32 pointings was observed, with pointing
centers separated by a primary beam FWHM as indicated in Figure 1. Both
crowded and empty regions were sampled. Given the low velocity
dispersion of the cluster, a 3.125 MHz bandwidth was sufficient to cover
the entire velocity range with a resolution of 10.3
km$\cdot$s$^{\mbox{\tiny -1}}$ after Hanning smoothing. A 70 minutes
integration time per pointing resulted in a rms noise of 0.79
mJy$\cdot$beam$^{\mbox{\tiny -1}}$ which corresponds to a minimum
detectable HI mass of 5$\times$10$^6$M$_\odot$ at 6-sigma at the beam
centers at a resolution of 45$^{\prime\prime}$ by 10
km$\cdot$s$^{\mbox{\tiny -1}}$. The data were flagged and calibrated in
AIPS, mosaiced and cleaned in MIRIAD and further analyzed in GIPSY.
Within the mosaiced images about 16\% of the entire cluster volume was
sampled. Currently, the full North-South and half of the East-West strip
is processed.

\section{Volume corrections}

Given the depth of the cluster and the primary beam attenuation of the
VLA, galaxies with small HI masses and broad profiles can not be
detected throughout the entire survey volume. Therefore, a simple
1/V$_{\rm max}$ volume correction was applied where V$_{\rm max}$ is the
volume in which a dwarf galaxy with a particular HI mass M$_{\rm HI}$
and line width W$_{20}$ could have been detected.
In an analytical approach, one can for each HI mass infer a line width
according to an empirical HI mass-line width relation:
W$_{20}$=0.16$\times$M$_{\rm HI}^{1/3}$ where W$_{20}$ is the line width
at 20\% of peak flux $F_{\rm peak}$. W$_{20}$ relates to the FWHM
$\Delta$V of a Gaussian profile according to
W$_{20}$=$\Delta$V$\cdot\sqrt{{\rm ln}5/{\rm ln}2}$. For a Gaussian
profile, $\int{F{\rm dv}}$ = $F_{\rm
peak}\cdot$W$_{20}$$\cdot\sqrt{\pi/4{\rm ln}5}$ where $F_{\rm
peak}$=6$\cdot\sigma_{\rm R}$ required for a detection and $\sigma_{\rm
R}$ is the noise at velocity resolution R.

Detectability is maximized under optimal smoothing conditions
(R=$\Delta$V) and the noise $\sigma_{\rm R}$ which corresponds to a
situation of optimal velocity smoothing is calculated according
$\sigma_{\rm R}$~=~$\sigma_{\rm R=10.3}\cdot\sqrt{10.3/{\rm R}}$ in
which $\sigma_{\rm R=10.3}$ is the noise in the highest velocity
resolution data.
Under this formalism, a maximum distance D$_{\rm max}$ at which an HI
mass M$_{\rm HI}$ can be detected is calculated according to 
$$ {\rm M}_{\rm HI} = 2.36\times10^5\cdot{\rm D}^2_{\rm max}\cdot\sigma_{\rm
R=10.3}\cdot\sqrt{{\rm W}_{20}}\cdot\sqrt{\pi 10.3/4{\rm ln}5}\cdot({\rm
ln}5/{\rm ln}2)^{1/4} $$

Of course, D$_{\rm max}$ varies across the primary beam due to its
attenuation characteristics which makes $\sigma_{\rm R}$ position
dependent. Considering the near and far side of the cluster volume,
V$_{\rm max}$ can easily be calculated from D$_{\rm max}$ and
$\sigma_{\rm R}$. This analytical approach is used to draw the solid
line in Figure 2. Ultimately, 1/V$_{\rm max}$ was calculated and applied
for each individual galaxy using its measured HI mass and line width.

\section{Results}

\begin{figure}
\plotone{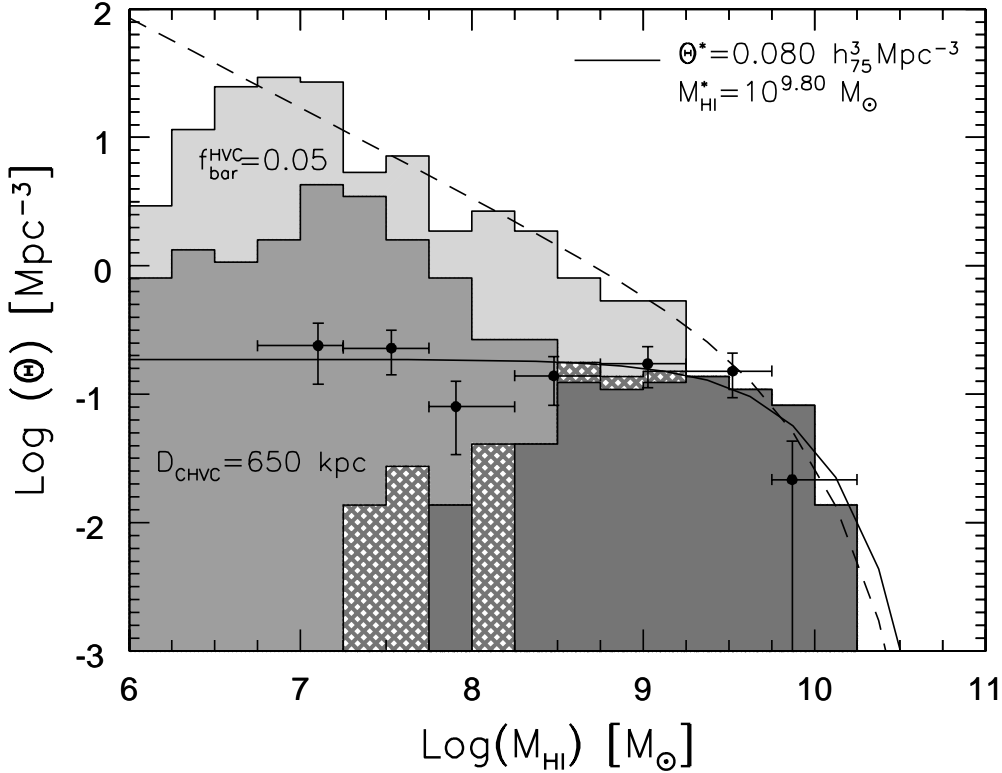}
 \caption{The Ursa Major HIMF in perspective. The volume corrected HIMF
in Ursa Major as measured by the VLA is plotted as points with errorbars
as explained in Figure 2. The dark-gray histogram shows the optically
selected HIMF to which the VLA HIMF was normalized at the high mass end.
The cross-hatched part shows contributions from galaxies intrinsically
fainter than the SMC. The light-gray histogram shows the expected
distribution of the UMa analogs of the proposed Local Group HVCs if they
would have similar space densities in UMa as in the Local Group. The
mass distribution is based on the assumption that the HVCs are
extragalactic and virialized with a dark matter fraction of 95\%. They
seem to predict a HIMF slope of $-$1.7 in UMa (dashed line). The
middle-gray histrogram is a similar expected distribution of UMa analogs
of the Compact HVCs identified by Braun \& Burton (1999) in case these
too are Local Group objects at a common distance of 650 kpc.}
 \end{figure}

The data cubes were smoothed in velocity to resolutions R of 20, 30, 40,
60 and 80 km$\cdot$s$^{\mbox{\tiny -1}}$ and at each velocity
resolution, the noise was determined and a 6-sigma clip was applied to
find the HI emission. In total 32 HI detections were made, all of them
have an optical counterpart in the CFHT images. All 19 previously known
cluster members within the imaged areas were detected, including all the
S0 systems. Only 13 new objects were discovered in 16\% of the UMa
volume. 

After applying volume corrections for the individual galaxies, the HIMF
in Ursa Major as measured by the VLA is plotted in the upper panel of
Figure 2. The analytical and idealized relation between M$_{\rm HI}$ and
V$_{\rm max}$ is plotted as a solid line in the upper panel which shows
that galaxies with M$_{\rm HI}$$>$10$^{8.75}$M$_\odot$ can be detected
throughout the entire imaged volume while 1/V$_{\rm max}$ becomes very
large for galaxies with M$_{\rm HI}$$<$10$^7$M$_\odot$.

The HI mass function in Ursa Major turns out to be nearly flat with 4
galaxies in the lowest HI mass bin. In accordance with the HIMF, the LF
from the CFHT data is also nearly flat ($\alpha=-1.1$) down to M(R)$=-10$.

Figure 3 shows the VLA derived HIMF in comparison with the optically
selected HIMF and the expections for HVCs in the Ursa Major volume. The
dark-gray histogram shows the previously known, optically selected UMa
HIMF to which the VLA HIMF was normalized at the high mass end.

The light-gray histogram shows the expected distribution of HI masses of
the population of Local Group HVCs, under the assumptions that they are
indeed of an extragalactic nature and virialized with a dark matter
fraction of 95\%, and that Ursa Major would have the same space
density of HVCs as the Local Group for which a volume of 15 Mpc$^3$ was
adopted. Such a population of extragalactic HVCs would give rise to a
slope of $-$1.7 as indicated by the dashed line which is highly
inconsistent with the actually measured data points.

The middle-gray histogram shows the expected distribution of HI masses
of the recently identified population of compact HVCs (Braun \& Burton
1999) in case they were located in the Local Group at a common distance
of 650 kpc, and again, if the space density in UMa would be the same as
in the Local Group. 

In conclusion, although this is a preliminary analysis of 85\% of the
data, the HI mass function in the Ursa Major cluster is flat down to HI
masses of 10$^7$ M$_\odot$ and a slope as steep as $-$1.7 is certainly
ruled out in UMa; it would have predicted 30 times more
10$^{7.5}$M$_\odot$ objects (well above the detection limit) than
observed with the VLA. No free-floating HI clouds were detected nor the
UMa analogs of the hypothesized Local Group extragalactic HVCs. Their
space density in UMa would have to be a factor $\sim$100 smaller than in
the Local Group.

\acknowledgements

The Very Large Array is a facility of the National Radio Astronomy
Observatory, a facility of the National Science Foundation operated
under cooperative agreement by Associated Universities Inc.

\end{document}